%% file: hadron2011.tex
\documentclass[a4paper,11pt]{article}

\usepackage{contribution}

\input{econfmacros}
\input{contributionmacros}

\begin{document}

\input{contribution}

\end{document}

%% file: econfmacros.tex
\newcommand{\weblink}[2][]{%
    \ifthenelse{\equal{#1}{}}%
    {\textnormal{\url{#2}}}%
    {\textnormal{\href{#2}{#1}}}%
}

\newcommand{\acknowledgements}[1]{%
  \bigskip\bigskip
  \textsf{\textbf{\Large Acknowledgements}} \\[2ex]
  {#1}
  \bigskip
}

\def\beq{\begin{equation}}
\def\eeq#1{\label{#1}\end{equation}}
\def\eeqn{\end{equation}}

\def\beqa{\begin{eqnarray}}
\def\eeqa#1{\label{#1}\end{eqnarray}}
\def\eeqan{\end{eqnarray}}

\let\bar=\overbar

\def\etal{{\it et al.}}

\def\Dslash{\not{\hbox{\kern-4pt $D$}}}
\def\dslash{\not{\hbox{\kern-2pt $\del$}}}

\def\msb{{\bar{\ssstyle M \kern -1pt S}}}

%% file: contributionmacros.tex
\newcommand{\contribution}[7][]{%
  \clearpage
  \thispagestyle{plain}
  \ifthenelse{\equal{#1}{}}
  {\hypersetup{pdftitle={#2}}}
  {\hypersetup{pdftitle={#1}}}
  \hypersetup{pdfauthor={{#3} {#4}}}
  {\centering\normalfont\LARGE\bfseries\sffamily #2 \par\nobreak}
  \lhead{}
  \chead{%
    \textit{\footnotesize XIV International Conference on Hadron Spectroscopy
      (\weblink[\textit{hadron2011}]{http://www.hadron2011.de}), 13-17 June 2011, Munich, Germany}%
  }
  \rhead{}
  \bigskip
  \begin{center}
    {#3} {#4}\ifthenelse{\equal{#6}{}}{}{\footnote{\weblink[#6]{mailto:#6}}}
    \ifthenelse{\equal{#7}{}}{}{#7} \\
    \textit{#5}
  \end{center}
  \bigskip
}

\renewcommand{\abstract}[1]{%
  \begin{center}
    \begin{minipage}{0.85\textwidth}
      \begin{footnotesize}
        #1
      \end{footnotesize}
    \end{minipage}
  \end{center}
  \bigskip
}

%% file: contribution.tex
%
%
%
%
%
{  


%

\contribution[Study of charmonium spectroscopy at BESIII]  
{Study of charmonium spectroscopy at BESIII}  
{Liangliang}{Wang}  
{Institute of High Energy Physics, Chinese Academy of Sciences\\
  Beijing, CHINA}  
{llwang@ihep.ac.cn}  
{on behalf of the BESIII Collaboration}  
%

\abstract{%
In this talk, we will present the results on the charmonium spin singlet states below the open charm threshold,
  including $h_c$, $\eta_c$, and $\eta_c(2S)$. The masses, widthes, and production rates of these states will be reported.
  The results are based on a data sample of 106 million $\psi^\prime$ events collected with the BESIII experiments at the BEPCII collider.
}
%

\section{Introduction}

In 2009, $(106\pm4)\times 10^6$ $\psi'$ events were collected with
BESIII detector at the upgraded BEPC (BEPCII) \cite{bes3-1}. All the
resent results on charmonium spectroscopy reported in this
proceeding are based on this set of data.

\section{Observation of $h_c$}
In 2010, the results on the production and decay of the $h_c$ at the
$\psi'$ resonance was reported by BESIII~\cite{hc-inc-bes3}, where
the distributions of mass recoiling against a detected $\pi^0$ were
studied to measure $\psi'\to\pi^0h_c$ both inclusively (E1-untagged)
and in events tagged as $h_c\to\gamma\eta_c$ (E1-tagged) by
detection of the E1 transition photon. In 2011, 16 specific decay
processes of $\eta_c$ in the decay mode of $h_c\to\gamma\eta_c$ are
studied to do the measurements of the $h_c$ properties in addition.
The simultaneous fit of the 16 $\pi^0$ recoil-mass spectra
(Figure~\ref{fig:hc_simFit}) yields
$M(h_c)=3525.31\pm0.11\pm0.15\mathrm{MeV/c^2}$ and
$\Gamma(h_c)=0.70\pm0.28\pm0.25\mathrm{MeV/c^2}$, where the first
errors are statistical and the second systematic. These preliminary
results are consistent with the previous BESIII inclusive results
and CLEOc exclusive results
($M(h_c)=3525.21\pm0.27\pm0.14\mathrm{MeV/c^2}$)~\cite{hc-exc-cleo}.

\begin{figure}[htb]
  \begin{center}
    \includegraphics[width=0.5\textwidth]{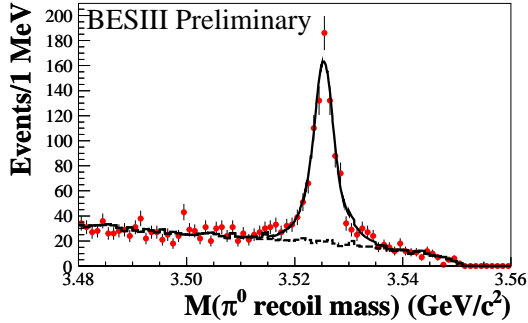}
    \caption{The summed $\pi^0$ recoil-mass spectrum of 16 specific decay
processes of $\eta_c$ in the decay mode of $h_c\to\gamma\eta_c$,
where the line is the fit result.}
    \label{fig:hc_simFit}
  \end{center}
\end{figure}

\section{Precision measurement of the $\eta_c$ properties}
With the largest $\psi'$ sample collected by BESIII, the $\eta_c$
mass and width are measured in the radiative transition
$\psi'\to\gamma\eta_c$, where six decay modes of $\eta_c$ are
involved: $K_S^0K\pi$, $K^+K^-\pi^0$, $\pi^+\pi^-\eta$,
$K_S^0K3\pi$, $K^+K^-\pi^+\pi^-\pi^0$ and $3(\pi^+\pi^-)$. A
simultaneous fit with the unique $\eta_c$ mass and width is
performed on the $\eta_c$ mass spectra, where the interference
between $\eta_c$ and non-$\eta_c$ decays is considered and the
quantum number of the non-$\eta_c$ components are assumed to be
$0^{-+}$. The corresponding interference phase angles in different
decay modes are found to be quite consistent and then set to the
same one in final fit. The mass spectra and the simultaneous fit for
different decay modes are shown in Figure~\ref{fig:etac_fit}. The
obtained results are $M(\eta_c)=2984.2\pm0.6\pm0.5\textrm{MeV/c2}$,
$\Gamma(\eta_c)=31.4\pm1.2\pm0.6\textrm{MeV}$, and
$\phi=2.41\pm0.06\pm0.04$rad, where the first errors are statistical
and the second systematic. The BESIII preliminary results are
consistent with those from two-photon
production~\cite{etac-gg-CLEO,etac-gg-Babar,etac-gg-Belle}, as well
as $J/\psi\to\gamma\eta_c$ by CLEOc~\cite{etac-Jpsi-Cleo}. And the
precision of the measured mass and width are improved.
\begin{figure}[htb]
  \begin{center}
    \includegraphics[width=0.8\textwidth]{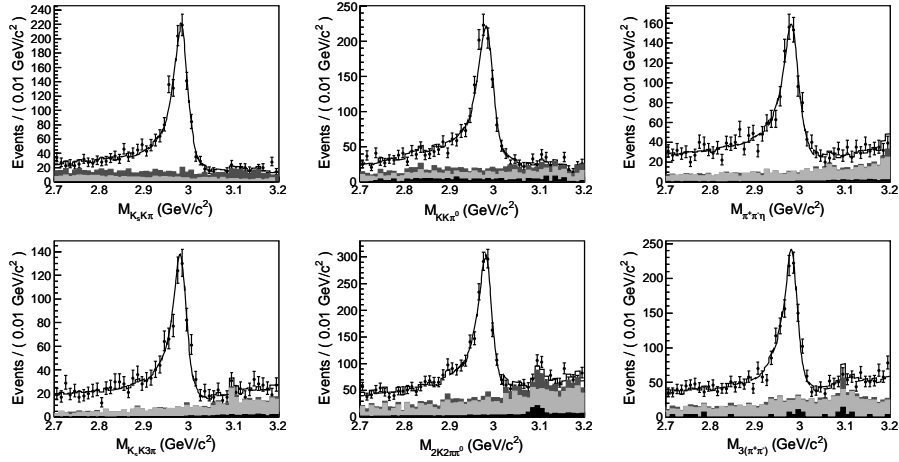}
    \caption{The mass spectra for different
decay modes, where the line is the result of the simultaneous fit.}
    \label{fig:etac_fit}
  \end{center}
\end{figure}

\section{The first observation of the M1 transition $\psi'\to\gamma\eta_c(2S)$}

BESIII observed this M1 transition $\psi'\to\gamma\eta_c(2S)$ with
the decay mode $\eta_c(2S)\to K_SK\pi$ for the first time.
Figure~\ref{fig:etacp} shows the preliminary result for the
invariant mass distribution of $K_S^0K\pi$ that the
three-constraints kinematic fit has been applied (where the energy
of the photon is allowed to be floating). The pure statistical
significance is more than $6\sigma$. The yielded events number is
$50.6\pm9.7$ and $M(\eta_c(2S))=3638.5\pm2.3\pm1.0\mathrm{MeV/c^2}$.
With the detection efficiency from MC simulation,
$B(\psi'\to\gamma\eta_c(2S)\to\gamma
K_S^0K\pi)=(2.98\pm0.57\pm0.48)\times10^{-6}$ is obtained. Combining
the result $B(\eta_c(2S)\to K\overline{K}\pi)=(1.9\pm0.4\pm1.1)\%$
from Babar, it is first calculated that
$B(\psi'\to\gamma\eta_c(2S))=(4.7\pm0.9\pm3.0)\times10^{-4}$ which
is consistent with the CLEOc's upper limit~\cite{etacp-cleo-c} and
prediction of potential model~\cite{MissChamoniumInB,HQproBcInRQM}
(the transition rate predicted by \cite{HQproBcInRQM} should be
$4.8\times10^{-4}$ if the mass of $\eta_c(2S)$ is updated to 3637MeV
according to PDG~\cite{PDG2011}), where the first errors are
statistical and the second systematic.

\begin{figure}[htb]
  \begin{center}
    \includegraphics[width=0.5\textwidth]{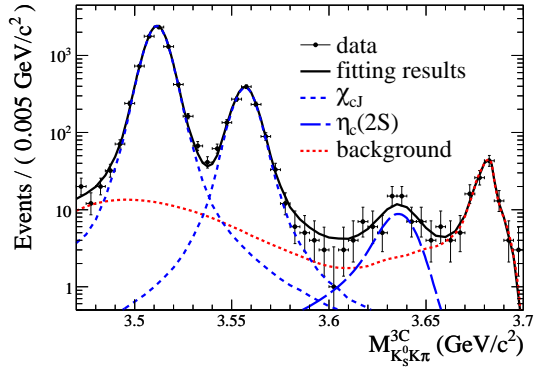}
    \caption{The invariant mass of $K_S^0K\pi$ from $\psi'\to\gamma K_S^0K\pi$.}
    \label{fig:etacp}
  \end{center}
\end{figure}

\section{Summary}

With the largest $\psi'$ data collected by BESIII, the following
results on Charmonium spectroscopy are obtained: the properties of
$h_c$ are measured with inclusive and exclusive methods
respectively; the properties of $\eta_c$ are precisely measured
using the radiative decays of $\psi'$, where the interference
between $\eta_c$ decays and non-$\eta_c$ decays is taken into
account; the M1 transition $\psi'\to\gamma\eta_c(2S)$ is observed
for the first time.

\acknowledgements{%
We thank the BEPCII group for excellent operation of the
accelerator, the IHEP computer group for valuable computing and
network support, and all the colleagues contribute on the physical
measurements.
}


%

}  


%% file: hadron2011.bbl
\begin{thebibliography}{99}

\bibitem{bes3-1} M. Ablikim {\em et al.} (BESIII Collaboration), {\it Phy. Rev. D} {\bf 81}, 052005 (2010).

\bibitem{hc-inc-bes3} M. Ablikim {\em et al.} (BESIII Collaboration), {\it Phys. Rev. Lett.} {\bf 104},
132002 (2010).

\bibitem{hc-exc-cleo} S. Dobbs {\etal} (CLEO Collaboration), {\it Phys. Rev. Lett.} {\bf 101},
182003 (2008).


\bibitem{etac-gg-CLEO} D. M. Asner {\etal} (CLEO Collaboration), {\it Phys. Rev. Lett.} {\bf 92},
142001 (2004).

\bibitem{etac-gg-Babar} B. Aubert {\etal} (BABAR Collaboration), {\it Phys. Rev. Lett.} {\bf 92},
142002 (2004).

\bibitem{etac-gg-Belle} S. Uehara {\etal} (Belle Collaboration), {\it Eur. Phys. J. C} {\bf 53}, 1
(2008).

\bibitem{etac-Jpsi-Cleo} R. E. Mitchell {\etal} (CLEO Collaboration), {\it Phys. Rev. Lett.} {\bf 102}, 011801 (2009).

\bibitem{etacp-cleo-c} D. Cronin-Hennessy, K. Y. Gao {\etal} (CLEOc Collaboration), {\it Phys. Rev. D} {\bf 81}, 052002 (2010).

\bibitem{MissChamoniumInB} Estia J. Eichten, Kenneth Lane, Chris Quigg, {\it Phys. Rev. Lett.} {\bf 89}, 162002 (2002).

\bibitem{HQproBcInRQM} D. Ebert, R. N. Faustov and V. O. Galkin {\it Phys. Rev. D.} {\bf 67}, 014027 (2003).

\bibitem{PDG2011} K. Nakamura {\etal} (Particle Data Group), {\it Journal of Physics G} {\bf 37},
075021 (2010) and 2011 partial update for the 2012 edition.

\end{thebibliography}
